\documentclass[iop]{emulateapj}
\usepackage{amssymb,amsmath} %\usepackage{natbib} %\usepackage{siunitx}
\usepackage[usenames,dvipsnames]{color}

\newcommand{\myemail}{lacerda.pedro@gmail.com}

\newcommand{\Red} {RC}
\newcommand{\Blue} {BC}

\newcommand{\add}[1]{{\color{Gray} #1}}

\slugcomment{\add{Accepted for publication in ApJL, 2012/10/8}} 

\shorttitle{Extinction in the Coma of Comet 17P/Holmes}
\shortauthors{Pedro Lacerda}

\begin{document}

\title{Extinction in the Coma of Comet 17P/Holmes}

\author{Pedro Lacerda}
\affil{Astrophysics Research Centre, Queen's University Belfast, Belfast
BT7 1NN, UK\\ {\tt \myemail}}
\and
\author{David Jewitt}
\affil{Department of Earth and Space Sciences and Department of
Physics and Astronomy, UCLA\\ 595 Charles Young Drive, Los Angeles, CA 90095-1567}

%%%%%%%%%%%%%%%%%%%%%%
\begin{abstract}

On 2007 October 29 the outbursting comet 17P/Holmes passed within
0.79\arcsec\ of a background star. We recorded the event using
optical, narrowband photometry and detect a 3\% to 4\% dip in stellar
brightness bracketing the time of closest approach to the comet
nucleus.  The detected dimming implies an optical depth
$\tau\approx0.04$ at 1.5\arcsec\ from the nucleus and an optical depth
towards the nucleus center $\tau_n<13.3$. At the time of our
observations, the coma was optically thick only within
$\rho\lesssim0.01\arcsec$ from the nucleus. By combining the measured
extinction and the scattered light from the coma we estimate a dust
red geometric albedo $p_d=0.006\pm0.002$ at $\alpha=16\degr$ phase
angle. Our measurements place the most stringent constraints on the
extinction optical depth of any cometary coma.

\end{abstract}


%%%%%%%%%%%%%%%%%%%%%%
\section{Introduction}

An extraordinary explosion in late 2007 brought comet 17P/Holmes
(hereafter, 17P) back to the limelight of scientific interest, more
than one hundred years after it was discovered by Edwin
\citet{1892Obs....15..441Holmes}.  The discovery and early interest
were triggered by an earlier, double outburst
\citep{1896ApJ.....3...41Barnard}. Since then and until the 2007
event, 17P had received little attention due to its rather ordinary
dynamical and photometric properties. 17P is a Jupiter-family comet,
with an orbit between Mars (perihelion $q=2.1$ AU) and Jupiter
(aphelion $Q=5.2$ AU) that has now experienced two explosive events
separated by a comparatively quiet period of over 100 years. Here we
report data taken shortly after the 2007 outburst to constrain the
properties of the dust coma of this comet.

Like most active comets, 17P appeared optically bright owing to
scattering from dust particles expelled from the nucleus by gas drag.
In principle, measurements of extinction from cometary dust particles
can be made by measuring the brightness of field stars when projected
behind the coma of a passing comet \citep{1983Icar...56..229Combes}.
Measurements of cometary extinction are potentially important as, when
combined with measurements of the scattered light, they can provide
direct estimates of the ensemble dust albedo.  In extreme cases,
extinction effects in cometary comae might influence the radiation
budget at the nucleus and so influence the mass loss rate caused by
sublimation.

Several such measurements have been reported
\citep{1984Icar...58..446Larson, 1984Icar...60..386Lecacheux,
1995Natur.373...46Elliot, 1999Icar..140..205Fernandez} but the results
are mostly negative or difficult to interpret, as a result of
uncertainties in the photometric data.
\citet{1999Icar..140..205Fernandez} reported an extinction of about
20\% in bright comet C/Hale-Bopp but, unfortunately, observed through
highly non-photometric skies that bring the significance of the
dimming into question.  Even deeper events were reported in 95P/Chiron
by \citet{1995Natur.373...46Elliot} and interpreted by them as
resulting from extinction in narrow, near nucleus dust jets.
Extinction from the diffuse coma of Chiron was not detected.

The bright coma of exploding comet 17P provided an opportunity to
search for measurable extinction of light from field stars.  The comet
impulsively ejected (2-90)$\times$10$^{10}$ kg of dust into the coma,
with a peak production rate 3.5$\times$10$^5$ kg/s on UT 2007 Oct
24.54$\pm$0.01 \citep{2011ApJ...728...31Li}.  These authors inferred,
from surface photometry alone, an optically thick region near the
nucleus subtending 0.1 arcseconds when the comet was at peak
brightness.  Extinction was reported by
\citet{2008A&A...479L..45Montalto} four days later and at much larger
angular distances from the nucleus (25 to 180 arcsec), where surface
photometry indicates immeasurably low line of sight optical depths
$<$2.5$\times$10$^{-3}$ \citep{2011ApJ...728...31Li}.

\begin{deluxetable}{ll}
  \tablecaption{Journal of Observations of 17P. \label{Table.Journal}}
   \tablewidth{0pt}
   %\tabletypesize{\tiny}
   \startdata
\hline
\hline
  UT Date                       & 2007 October 29 \\
  Time of closest approach      & UT 12:28:42 \\
  Heliocentric distance, $R$    & 2.455 AU\\
  Geocentric distance, $\Delta$ & 1.627 AU\\
  Phase angle, $\alpha$         & 15.6\degr\\
  Rate of motion 		& 26.4\arcsec\ hr$^{-1}$ \\
  Weather                       & Photometric \\
  Telescope                     & UH 2.2 m\\
  Instrument                    & TEK \\
  Pixel scale                   & 0.219\arcsec/pixel \\
  Seeing                        & $0.9\pm0.1$\arcsec \\
  Filters (Exp. Time)           & \Blue\ (120 s), \Red\ (25 s), $R$
(1 s) 
   \enddata
\end{deluxetable}

\begin{figure*} 
  \centering
  \includegraphics[width=0.75\textwidth]{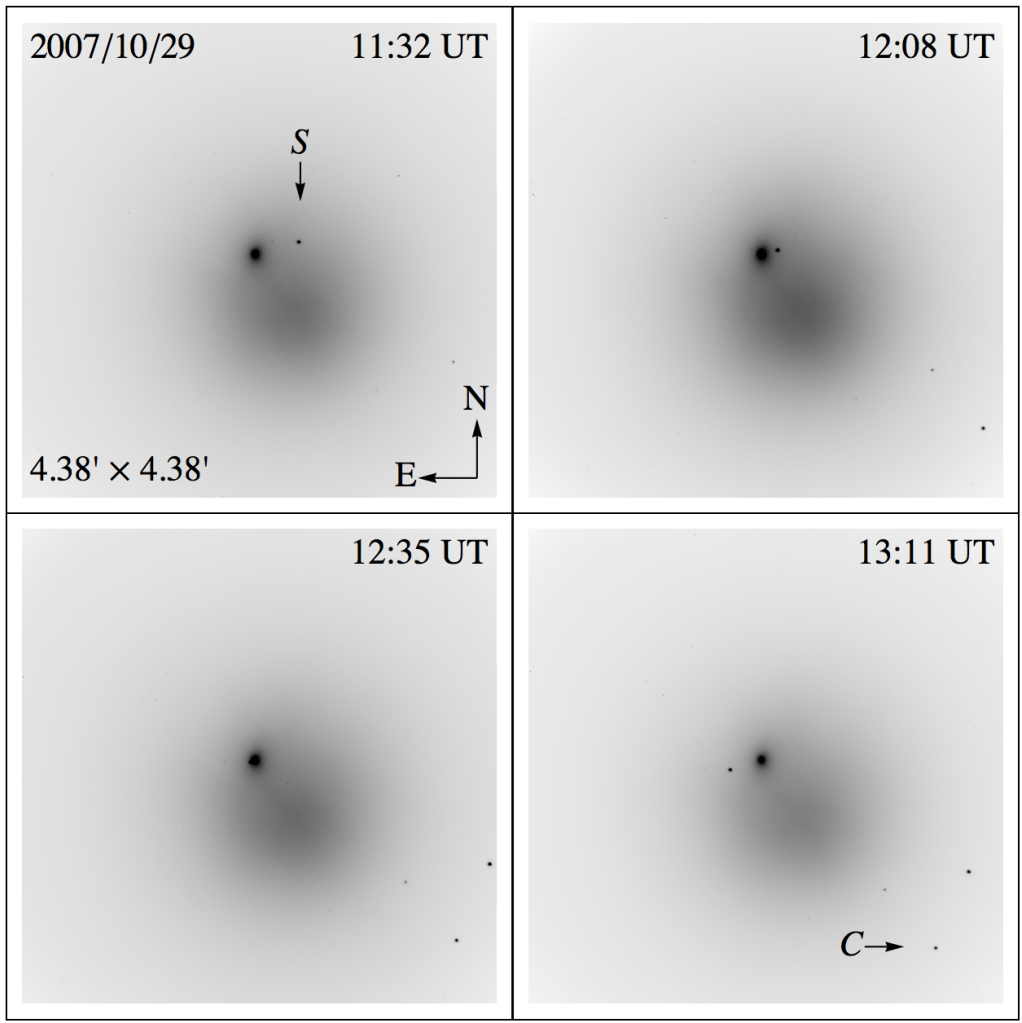}

  \caption[f1] {\Red-band snapshots of the stellar appulse of 17P and
  star S. Differential photometry between stars S and C revealed
  dimming of the former as it passed behind the central coma of 17P.
  These are sub-frame sections of the TEK 2k CCD.}

  \label{Fig.StarPath}
\end{figure*}

%%%%%%%%%%%%%%%%%%%%%%
\section{Observations} \label{Sec.Observations}

Observations of 17P were obtained using the University of Hawaii 2.2-m
telescope situated atop Mauna Kea. The telescope was equipped with a
Tektronix TEK charged coupled device (CCD), which holds
2048$\times$2048 pixels each 0.219\arcsec$\times$0.219\arcsec\ on the
sky, and is read out in $\sim$35 s. Our measurements were obtained
through the Hale-Bopp narrowband blue continuum (\Blue; $\lambda=4453$
\AA, $\Delta\lambda=61$ \AA\ FWHM) and red continuum (\Red;
$\lambda=7133$ \AA, $\Delta\lambda=58$ \AA\ FWHM) filters
\citep{2000Icar..147..180Farnham,2004AJ....128.3061Jewitt}. For
absolute calibration purposes we obtained a number of broadband $R$
(Kron-Cousins-type filter) images of 17P and Landolt standard stars
throughout the night. To avoid saturation of the bright central region
we were forced to use an integration time of 1 s, shorter than the
minimum advised 5 s for the TEK camera. Integrations shorter than
$\sim$5 s suffer from a systematic spatial pattern due to the finite
shutter time that is more noticeable towards the edges of the CCD.
Using 1 s dome flatfield images we found that within our region of
interest (near the center of the CCD as we tried to image the full
circular coma) the shutter pattern uncertainty amounts to $<1\%$. The
data were calibrated using bias frames and flatfield images obtained
from dithered, median-combined images of the twilight sky.  The night
was photometric with very stable seeing, $0.9\pm0.1$\arcsec.  We set
the telescope to track the non-sidereal rate (26\arcsec\ hr$^{-1}$) of
17P and, as a result, stellar trailing in the longer \Blue\ exposures
is comparable to the seeing.  A journal of the observations can be
found in Table\ \ref{Table.Journal}.

%%%%%%%%%%%%%%%%%%%%
\section{Photometry} \label{Sec.Photometry}

We identified several field stars in the projected path of the comet
and selected the one (TYC 3334-166-1) having the highest brightness
($R=11.0\pm0.1$ mag) and minimum impact parameter ($\rho=0.79\arcsec$)
for this study; we refer to this as star S. Figure\ \ref{Fig.StarPath}
shows some representative images in \Red\ band.  By 2007 October 29,
the coma of 17P had expanded to fill the entire 7.5\arcmin\ field of
view of the TEK detector \citep{2010MNRAS.407.1784Hsieh}.
Consequently, each CCD pixel included contributions from the
background sky and from the 17P coma.  To measure small changes in the
flux from star S it was necessary to subtract the contribution from
the sky and the coma of 17P. We attempted this subtraction in two
different ways. 

\begin{figure} 
  \centering
  \includegraphics[width=\columnwidth]{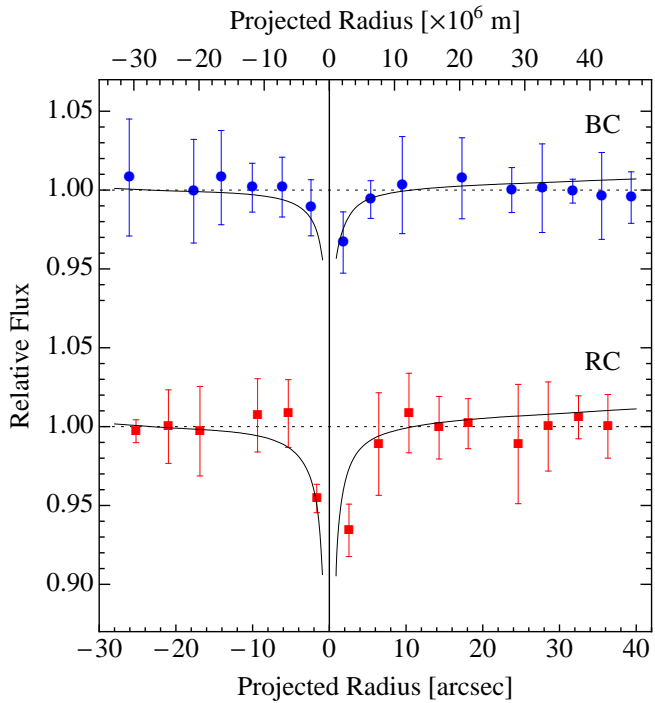}

  \caption[f2] {Lightcurve of star S versus projected
    radius from the nucleus of 17P through filters \Blue\ (blue circles)
    and \Red\ (red squares). Negative projected radii indicate approach.
    The flux of star S was measured relative to that of star C.
    Overplotted as a solid black line is the inverted coma brightness
    profile, vertically shifted and scaled to fit the lightcurve of star
  S.}

  \label{Fig.FluxVsDistance}
\end{figure}

\begin{figure} 
  \centering
  \includegraphics[width=\columnwidth]{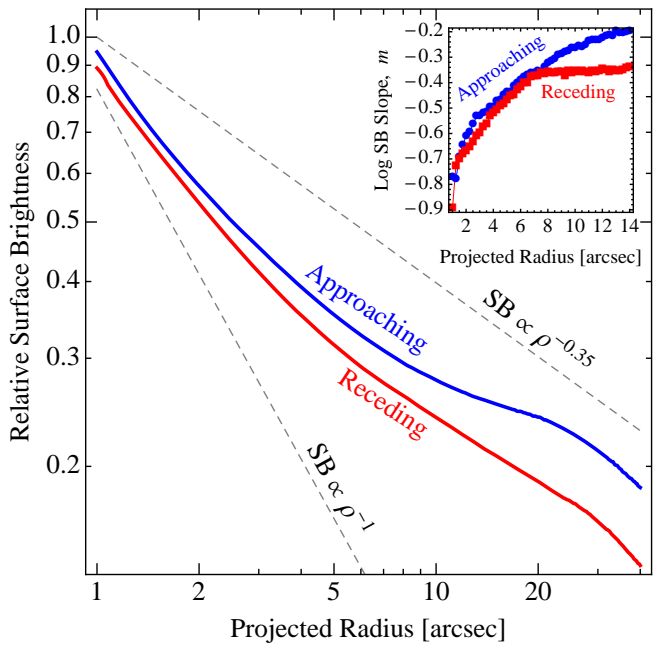}

  \caption[f3] {Surface brightness of the coma of 17P
    versus projected distance from the nucleus along the path of star
    S. Overplotted as dashed lines are slopes proportional to
    $\rho^m$, for $m=-0.35$ and $m=-1$. The inset shows the slope of the
  profile as a function of projected distance.}

  \label{Fig.ComaSBProfile}
\end{figure}

Firstly, we employed pair-wise subtraction of near-consecutive frames
aligned on the comet nucleus. Before subtraction, the frames were
scaled by dividing each by the median pixel value measured in a
0.7\arcmin$\times$0.7\arcmin\ square region centered on the comet
nucleus. The point spread function (PSF) of the star overlapped
slightly in consecutive frames (17P moved $\sim$3.9\arcsec\ between
frames) so we used every second frame for pair subtraction. A second
method consisted of constructing ``starless'' frames by
median-combining sets of three near-consecutive exposures.  Each
median frame was then subtracted from the central of the medianed
exposures.  As before, the median of three consecutive exposures
retained artifacts due to the star, so we used sequences 1-3-5 to
construct the median to be subtracted from frame 3. Medians of 5
frames (median of frames 1-3-5-7-9 subtracted from frame 5) produced
similar results to 3-frame medians.  Neither of the methods was ideal
due to the fast-expanding coma, to slight changes in the seeing, and
to imperfect alignment of the nucleus in different frames. Both
methods left traces of the coma that were strongest near the nucleus,
but that amounted to at most, 1\% of the flux from star S. The
following results are based on the pair-wise subtraction method which
we found to produce slightly better (less noisy) results.

\begin{deluxetable}{ccccc}
\tablecaption{Model Optical Depths for 17P. \label{Table.Tau}}
\tablewidth{0pt}
%\tabletypesize{\tiny}
\tablehead{
	\colhead{$n(r)$ model} & $q$ & $q(\rho<1\arcsec)$ & $\tau(1.5\arcsec)$ & $\tau_n$ 
}
\startdata
   $n(r)\propto r^q$ & $-2$     & $-2$           & 0.04 & 13.3 \\
   $n(r)\propto r^q$ & $-1.35$  & $-1.35$        & 0.04 & 0.19 \\
   from SB           & variable & $q(1\arcsec)$  & 0.04 & 3.46 \\
   from SB           & variable & $-2$           & 0.04 & 8.28 \\
   \enddata
\tablecomments{Columns are (1) model for $n(r)$ where ``from SB'' indicates
  that $n(r)$ is inferred from coma surface brightness profile, (2)
  assumed $q$ if constant, (3) assumed $q$ for coma inner to 1\arcsec,
  not resolved by our photometry, (4) measured optical depth at
1.5\arcsec\ impact parameter, (6) model optical depth towards
nucleus.}
   \end{deluxetable}

We performed circular aperture photometry centered on star S in each
of the sky- and coma-subtracted images. Tests with increasing
apertures indicated that an aperture radius of 3.3\arcsec\ produced
the most stable photometry when star S projected far from the nucleus.
Because of the subtraction, the flux surrounding the star should be
zero, with a combined uncertainty that is approximately Gaussian.
However, we chose to subtract the sky measured in an annulus
surrounding the star to remove local residual subtraction offset. To
estimate the uncertainty in the flux from star S at position $i$ we
used the standard deviation of the fluxes measured at that same
position $i$ in the $N-1$ frames ($N=15$) for which the star is not
present in the aperture. In addition, to confer protection against
small seeing variations and other time-dependent variations we
employed relative photometry. For this, we used another star (C, see
Fig.\ \ref{Fig.StarPath}) that lies further ($>$2.3\arcmin) from the
nucleus, which we measured in the same way as star S.  Star C
($R\sim13$ mag) lies distant enough from the bright central region of
the coma that it should suffer much smaller extinction than star S
\citep{2011ApJ...728...31Li}. We computed the ratio of the fluxes of
stars S and C as a function of time. 

Figure\ \ref{Fig.FluxVsDistance} shows the relative flux of stars S
and C versus projected radius from the nucleus through the \Blue\ and
\Red\ filters. We divided the curve by its median value so that it
explicitely represents the fractional dimming of star S. Near closest
approach to the nucleus of 17P, the star dims by about 3\% in \Blue\
(at 1.7\arcsec\ from the nucleus) and about 4\% in \Red\ (at
1.5\arcsec). The nearest \Blue\ and \Red\ measurements occur on
opposite sides of the nucleus and, for both bands, the brightness dip
is visible in two separate measurements bracketing the nearest
projected distance from 17P. The two \Blue\ points are 0.6$\sigma$ and
1.7$\sigma$ below the median, while the two \Red\ points are
5.1$\sigma$ and 4.0$\sigma$ below the median.  Assuming that each pair
of measurements is uncorrelated (i.e., their probabilities can be
multiplied) the combined significance of the two dips is 2.0$\sigma$
in \Blue\ and 6.6$\sigma$ in \Red. We note that both bands show a
consistent trend of stronger dimming on the post-appulse side of the
nucleus. This is true even though for \Red\ the closest post-appulse
measurement lies slightly farther from the nucleus (at
$\sim2.5\arcsec$) than the nearest pre-appulse point.

The observed dimming can be converted to an optical depth $\tau$ using
\begin{equation} \tau(\rho)=-\ln[F(\rho)/F_0] \label{Eq.Tau}
\end{equation} where $\rho$ is the impact parameter at which the
dimming is detected, $F(\rho)$ is the star S flux at $\rho$ and $F_0$
is the unimpeded star S flux.  Substituting 0.97 and 0.96 into
equation (\ref{Eq.Tau}) we obtain optical depths
$\tau(1.7\arcsec)=0.03$ in \Blue\ and $\tau(1.5\arcsec)=0.04$ in \Red.
In the remainder of the paper we use the higher significance \Red\
optical depth.

%%%%%%%%%%%%%%%%%%%%%%
\section{Discussion} \label{Sec.Discussion}

%%%%%%%%%%%%%%%%%%%%%%
\subsection{The optical depth to the nucleus} \label{Sec.NucleusTau}

Assuming a spherically symmetric coma composed of dust particles of
radius $a$ with a radial number density dependence $n(r)$, the optical
depth along a line-of-sight to the center of the nucleus, $\tau_n$,
can be related to the optical depth at a projected distance $\rho$ by
\begin{equation}
\frac{\tau_n}{\tau(\rho)}=\frac{\left(1/2\right)\int_{r_n}^{\infty}\pi
a^2\, n(r)\mathrm{d}r}{\int_{0}^{\infty}\pi a^2\, n(r')\mathrm{d}l}
\label{Eq.TauNucTauRho} \end{equation} where $r'=\sqrt{l^2+\rho^2}$,
$l$ is measured along the line-of-sight to projected radius $\rho$,
$r_n$ is the radius of the nucleus, and the $(1/2)$ factor accounts
for the opaqueness of the nucleus. 
%If we take a dust number density of the form $n(r)\propto r^q$ then
%the integrals in equation (\ref{Eq.TauNucTauRho}) can be solved
%analytically to give \begin{equation} \tau_n=0.5
%\left(r_n/\rho\right)^{q+1} \tau(\rho). \label{Eq.SimpleTauNuc}
%\end{equation}
%
We can use the observed coma surface brightness profile (surface
brightness versus projected radius from the nucleus, $\rho$) to infer
the dust number density profile, $n(r)$.  If we assume the latter
varies proportionally to $r^q$ then by integration the former will
vary proportionally to $\rho^m$, where $m=q+1$.  Figure
\ref{Fig.ComaSBProfile} shows the near-nucleus surface brightness
profile from our \Red\ data measured on a 3\arcsec-wide band centered
on the path of star S across the coma.  The profile is steepest near
the nucleus, where it approaches the canonical slope $\rho^{-1}$ of a
spherically symmetric, steady state coma, and is on average
$\propto\rho^{-0.35}$ in the inner 30\arcsec\ of the coma.  Using
$n(r)\propto r^q$ with the maximum slope $q=-2$ (from $m=-1$), and
replacing $\tau(1.5\arcsec)=0.04$ and $r_n=1.7$ km
\citep{2009A&A...508.1045Lamy} into equation (\ref{Eq.TauNucTauRho})
we obtain $\tau_n=13.3$.  If, instead, we use the average inner coma
slope $q=-1.35$ we obtain $\tau_n=0.19$ (Table\ \ref{Table.Tau}). The
former, larger estimate can be regarded as an upper limit to the
optical depth towards the nucleus.  Steeper slopes than $m=-1$ are
generally due to solar radiation pressure which has insufficient time
to act in the inner 1\arcsec\ ($\sim$1000 km) of the coma
\citep{1987ApJ...317..992Jewitt}.

Shown as an inset in Fig.\ \ref{Fig.ComaSBProfile} is the measured
surface brightness profile slope as a function of projected radius
$\rho$, which varies from $m\sim-0.8$ at $\rho<3\arcsec$ to
$m\sim-0.3$ at $\rho\sim12\arcsec$. These slopes match the $m\sim-0.3$
to $-0.2$ found by \citet{2012arXiv1209.2165Stevenson} at distances
10000 to 20000 km (8\arcsec\ to 16\arcsec\ in our data) from the
nucleus in data obtained a few days to a few weeks after our own
observations. We can use the measured radial dependence of $m$ to
solve numerically the integrals in equation (\ref{Eq.TauNucTauRho})
and improve our estimate of the optical depth to the nucleus,
$\tau_n$. Table\ \ref{Table.Tau} summarizes our results where, again,
we used $\tau(1.5\arcsec)=0.04$ and $r_n=1.7$ km.  Because we only
measure $m$ to within approximately $1\arcsec$ of the nucleus we need
to assume the value of $q$ in the innermost 1\arcsec\ of the coma.  We
consider two possibilities: the limiting case $q(\rho<1\arcsec)=-2$,
and a constant $q(\rho<1\arcsec)=q(1\arcsec)$. In both cases we obtain
optical depths to the nucleus only slightly larger than unity, falling
off to $\tau=1$ at $\rho\sim0.01\arcsec$. 

Our measurement is consistent with the findings of
\citet{2011ApJ...728...31Li} who used coarser spatial resolution
photometry on data taken around 2012 October 24.5 to conclude that the
coma of 17P was optically thick only within a tiny central region,
$\sim0.1\arcsec$. They find a peak optical depth towards the nucleus
$\tau_n\sim65$ on October 24.8. Another detection of extinction by the
coma of 17P was reported by \citet{2008A&A...479L..45Montalto}, using
an ensemble of 20 stars located between 25\arcsec\ and 180\arcsec\
from the nucleus of the comet. They infer optical depths
$0.3<\tau<0.5$, two to three orders of magnitude larger than the
$1\times10^{-4}<\tau<4\times10^{-3}$ implied by our measurements
in the region $25\arcsec<\rho<180\arcsec$.

%%%%%%%%%%%%%%%%%%%%%%
\subsection{The coma dust albedo}

The extinction obtained in \S\ref{Sec.NucleusTau} can be combined with
the measured scattered light, to provide a direct estimate of the coma
dust albedo. We here use the $R$ band images mentioned in
\S\ref{Sec.Observations}. Assuming that the inner coma of 17P is
spherically symmetric then the optical depth at projected radius
$\rho$ from the nucleus can be written \begin{equation} \tau(\rho)
\approx
  \frac{A_d}{\pi \left(2\,\rho\,\delta\rho \right)\Delta^2}
  \label{Eq.CDA1} 
\end{equation} where $\pi\left(2\,\rho\,\delta\rho\right)\Delta^2$ is
the projected area of the annulus centered on the nucleus with inner
radius $\rho-\delta\rho/2$ and outer radius $\rho+\delta\rho/2$, and
$A_d$ is the total effective scattering cross-section of the dust
grains within the annulus.  The latter can be calculated from
\citep{1916ApJ....43..173Russell} \begin{equation}\label{Eq.CDA2}
  p_d\,A_d=2.24\times10^{22}\,\pi\,10^{0.4\left[m_\sun-m_d(\rho)\right]}
\end{equation} where $p_d$ is the red albedo of the dust grains,
$m_{\sun}=-27.11$ is the apparent red magnitude of the Sun, and
$m_d(\rho)$ is the absolute red magnitude\footnote{The absolute
magnitude is the apparent magnitude reduced to unit heliocentric and
geocentric distances and to zero degrees phase angle.} of the dust
contained within the annulus defined above. Equations (\ref{Eq.CDA1})
and (\ref{Eq.CDA2}) can be combined to yield
\begin{equation}\label{Eq.Albedo}
  p_d=\frac{2.24\times10^{22}\,10^{0.4\left[m_\sun-m_d(\rho)\right]}}{2\,\rho\,\delta\rho\,\Delta^2\,\tau(\rho)}.
\end{equation} We take $\rho=1.5\arcsec$, $\delta\rho/2=0.219\arcsec$
(1 pixel), $\tau(1.5\arcsec)=0.04$ and calculate the magnitude of the
coma in the annulus using \begin{equation} \begin{split}
  m_d(1.5\arcsec) &=
  -2.5\log_{10}\left(10^{-0.4m_\mathrm{out}}-10^{-0.4m_\mathrm{in}}\right)
  \\ &= 7.17\pm0.47\,\mathrm{mag}
\end{split}
\end{equation} where $m_\mathrm{out}=6.04\pm0.11$ mag and
$m_\mathrm{in}=6.51\pm0.11$ mag are the absolute red magnitudes
measured within the outer and inner radii of the annulus. Replacing
those quantities into equation \ref{Eq.Albedo} we obtain
$p_d=(6\pm2)\times10^{-3}$. 

Our estimated dust albedo is significantly lower than an earlier,
independent measurement using a different technique by
\citet{2010ApJ...714.1324Ishiguro}.  Those authors used combined
optical and infrared observations of 17P taken on 2010 October 25-28
to find an albedo $0.03\leq p_d\leq 0.12$.  Both estimates were
obtained at similar phase angle, $\alpha=16\degr$.  Interestingly,
\citet{2010ApJ...714.1324Ishiguro} report a decrease in albedo with
time, from $p_d=0.12\pm0.04$ on October 25 to $p_d=0.03\pm0.01$ on
October 28.  Our estimate follows the trend but we find a 5$\times$
lower albedo just a day later, on October 29.

Earlier attempts to estimate the reflectivity of cometary dust grains
using stellar appulses have also found surprisingly low albedos.
\citet{1984Icar...58..446Larson} observed comet Bowell (1980b) and
found a dust albedo of $p_d\sim1.5\times10^{-3}$. They suggested that
photons might go through two reflections ($0.0015\approx0.04^2$)
before escaping the dust grain to reconcile their findings with the
typical 4\% albedo of cometary surfaces.
\citet{1999Icar..140..205Fernandez} reported an albedo
$p_d\approx0.01$ for the coma dust of comet Hale Bopp.

Such low albedos are not unprecedented. Spacecraft imaging of comet
Borrelly revealed very dark regions with geometric albedos $p<0.01$ at
wavelengths $0.5\,\mu\mathrm{m}<\lambda<1.0\,\mu\mathrm{m}$
\citep{2002Sci...296.1087Soderblom,2004Icar..167...37Nelson}.
High-porosity carbon-based aerogels are also known to have very low
albedos over a broad range of wavelengths
\citep{2001JNCS..285..210Merzbacher}. The least reflective carbon
aerogels (reflectivities as low as 0.003) are made of $\sim$1 nm
amorphous carbon particles and $\sim$10 nm voids resulting in bulk
porosities larger than 90\%. Indeed, theoretical studies predict that
the albedo of porous grain aggregates of small particles should
decrease with increasing bulk porosity \citep{1990ApJ...361..251Hage},
reaching sub-1\% reflectivities at void volume fraction of about 70\%
\citep{2004MNRAS.355..191Napier}.  The low reflectivity of carbon
aerogels is due in part to strong absorption from amorphous carbon but
results mostly from the rough, highly porous physical structure of the
material (silica aerogels are also highly porous but are smoother on 1
$\mu$m scales).  Our observations and those of others may be
suggesting that comet dust is highly porous and, in the case of 17P,
possibly carbon-rich. We note that a significant amount of amorphous
carbon is required to explain Spitzer spectra of 17P taken on 2007
November 10, only a few days after our observations
\citep{2010Icar..208..276Reach}.

%%%%%%%%%%%%%%%%%%%%%%
\section{Conclusions}

We used narrowband photometry of a background star that passed within
$<1\arcsec$ of the nucleus of outbursting comet 17P/Holmes to
constrain the extinction optical depth of the cometary dust coma. Our
conclusions are as follows:

\begin{itemize}

  \item We detect a 3 to $4\%$ dimming of the background star
    depending on wavelength that coincides with the nearest projected
    distance to the cometary nucleus. If caused by extinction due to
    coma dust, this dimming implies an optical depth $\tau=0.03$ to
    0.04 at projected radius 1.5\arcsec\ from the comet nucleus. We
    infer that the coma was optically thick only in a tiny region,
    $\sim0.01\arcsec$ in radius, surrounding the nucleus. Our
    measurements are compatible with those by
    \citet{2011ApJ...728...31Li} but difficult to reconcile with those
    by \citet{2008A&A...479L..45Montalto}.

  \item We combine the measured extinction with photometry of the coma
    to estimate a dust albedo of ($0.6\pm0.2$)\%. Our albedo estimate
    is 5$\times$ lower than an independent measurement using optical
    and thermal observations of light scattered by the dust
    \citep{2010ApJ...714.1324Ishiguro}, but it is consistent with the
    darkest regions on comet Borrelly, and with the dust being
    composed of very small, highly porous grains of carbonaceous
    composition.

\end{itemize}

%%%%%%%%%%%%%%%%%%%%%%
\section*{Acknowledgments}

We thank Yan Fern\'andez for a very helpful review, and Andrew McNeill
for useful contributions to this work.  We appreciate support from a
NASA Outer Planets Research grant to DJ.

%%%%%%%%%%%%%%%%%%%%%%
\bibliographystyle{apj}

\end{document}